Interface-mediated softening and deformation mechanics in amorphous/ amorphous nanolaminates

Vivek Devulapalli[1] *, Fedor F. Klimashin[1], Manuel Bärtschi[2], Stephan Waldner[3], Silvia Schwyn Thöny[3], Johann Michler[1, 4], Xavier Maeder[1] *

[1] EMPA, Laboratory for Mechanics of Materials & Nanostructures, Feuerwerkstrasse 39, 3602 Thun, Switzerland

[2] RhySearch, Optical Coating and Characterization Laboratory, Werdenbergstrasse 4, 9470, Switzerland

[3] Evatec Ltd., Hauptstrasse 1a, 9477 Truebbach, Switzerland

[4] EPFL, Institute of Materials, Lausanne 1015, Switzerland

*Corresponding authors' email: vivek.devulapalli@empa.ch, xavier.maeder@empa.ch

**Abstract**

Interfaces govern the unique mechanical response of amorphous multilayers. Here, we examine nanoindentation hardness and deformation behaviour of amorphous-amorphous $Ta_2O_5$/$SiO_2$ nanolaminates with bilayer thicknesses ranging from 2 nm to 334 nm. Whilst monolithic $SiO_2$ exhibits catastrophic failure through a single dominant shear band, multilayer architectures demonstrate varied deformation mechanisms. Hardness decreases with reduced bilayer thickness, from 7.7 GPa at 334 nm to 5.5 GPa at 2 nm spacing, contrasting with crystalline systems, which strengthen with decreasing spacing. Cross-sectional transmission electron microscopy reveals that fine bilayer spacings promote closely spaced vertical shear bands with bilayer compression, while coarser spacings show fewer, widely spaced shear bands with chemical intermixing. Scanning electron diffraction mapping demonstrates significant densification beneath indents. The high interface density facilitates strain accommodation that prevents catastrophic failure typical of brittle amorphous materials.

**Main manuscript**

Amorphous $Ta_2O_5$ owing to its higher packing density exhibits plasticity [1,2]. In contrast, most amorphous oxides like $Al_2O_3$ or $SiO_2$, while possessing exceptional strength and hardness, are limited by their propensity for highly localised shear band formation at room temperature, leading to catastrophic brittle failure and restricting engineering applications [3,4]. However, when the brittle oxides are confined to nanoscale dimensions or constrained within multilayered architectures, they can exhibit remarkable plasticity [5–9]. This discovery has led to intensive research into amorphous-amorphous nanolaminates (A/ANLs), comprising alternating amorphous layers with differing compositions. Such multilayers demonstrate enhanced mechanical as well as optical properties enabling applications across diverse fields, including structural engineering, microelectronics, optoelectronics, and biomedical devices [10–15]. Far less is understood when ductile and brittle amorphous oxides are combined, where plastic accommodation and shear localization may interact to produce different mechanical responses.

While interfaces in crystalline-amorphous multilayers are well-established as barriers to shear band propagation [9,16], the role of interfaces in A/ANLs remains contentious. Studies report conflicting interlayer thickness-dependent behaviour: some demonstrate strengthening with decreasing bilayer spacing due to interface obstruction of shear bands [12,13,15], while others show thickness-independent properties [17,18,10]. For example, experimental studies on ZrCu/ZrCuAl metallic glass based nanolaminates showed shear bands blocked by interfaces due to significant chemical intermixing [15]. In contrast, nanoindentation of CuNb-based and CuTa/CoZrNb systems demonstrated hardness values equivalent to rule-of-mixtures predictions regardless of bilayer period, suggesting interfaces offer no resistance to shear band propagation [18]. These contradictions highlight poor understanding of interface-mediated deformation in A/ANLs.

While metallic A/ANLs have been well studied, have received less attention, despite their distinct bonding and properties. $Ta_2O_5/SiO_2$ multilayers are particularly relevant for antireflective coating applications due to their excellent optical properties and environmental stability. Varying the bilayer thickness in such nanolaminates alters the optical band gap through quantum confinement, essential for tuning their absorption edges [19–22]. Understanding how bilayer thickness affects both optical and mechanical behaviour is critical for robust coating design. This study systematically examines the effect of interface density on mechanical response in $Ta_2O_5/SiO_2$ nanolaminates using nanoindentation. By investigating thicknesses from a few nanometres to hundreds of nanometres, we aim to elucidate the deformation mechanisms governing oxide A/ANLs.

All films were deposited in a Clusterline 200 BPM magnetron sputtering system (Evatec AG, Switzerland) using reactive magnetron sputtering with identical process parameters, varying only the substrate table rotation speed. Silicon and tantalum targets were operated at 4 kW and 7 kW pulsed DC power, respectively, with Ar gas distributed to both targets. The $O_2$ flow regulated via plasma emission monitoring which enables the deposition of fully oxidised layers. The resulting deposition rate was 0.766 nm/s. The rotating substrate table enabled continuous bilayer deposition, with thickness ratios controlled by power adjustment and overall bilayer thickness tuned by rotation speed. Pure $SiO_2$ film was deposited at the same sputter source using plasma emission monitoring, but an additional capacitively coupled plasma source was used to provide surface energy enhancement via accelerated $O_2$ ions. Seven multilayer systems, each 1μm thick, with bilayer thicknesses of 2 nm, 4 nm, 10 nm, 20 nm, 40 nm, 200 nm, and 334 nm were deposited at room temperature on single crystal Si substrates, along with two monolithic reference films. In all the multilayers, $Ta_2O_5$ was deposited first followed by $SiO_2$.

Nanoindentation (Zwick-Roell ZHN, 2N measuring head) was performed in quasi-continuous stiffness measurement mode (strain rate 0.05 $s^{-1}$, oscillation frequency 40 Hz) using a Berkovich tip (Synton-MDP Ltd., Switzerland). Penetration depth was limited to one-third of the film thickness. Depth-resolved hardness profiles were obtained following Oliver and Pharr analysis [23]. The film hardness was determined by averaging values from a penetration depth range of 50-250 nm (see Fig. S1), in accordance with the 10% of film thickness criterion [24,25]. Nine indentations per sample with >20 μm spacing ensured statistical reliability. Additional cube corner indents (Synton-MDP Ltd., Switzerland) probed crack growth.

Cross-sectional transmission electron microscopy (TEM) specimens were prepared using focused ion beam (FIB) (Tescan Lyra) with 30 kV $Ga^+$ ions followed by 5 kV cleaning. TEM analysis (Thermo Fischer Titan Themis 200) included bright-field, dark-field, high-resolution imaging and high angle annular dark field (HAADF) imaging. Scanning precession electron diffraction (SPED) was performed in the same TEM using a NanoMEGAS DigiSTAR system with a 2 nm step size to generate Four-Dimensional scanning TEM (4D-STEM) data.

Fig. 1 presents SEM micrographs of nanoindents on all specimens, revealing the varied deformation behaviour of each film. Fig. 1a shows pronounced pile-up around the indents in amorphous $Ta_2O_5$, without evidence of radial or lateral cracking. The observed pile-up suggests that deformation proceeds predominantly by shear flow, consistent with the reported ductile response of amorphous $Ta_2O_5$ [1,2,26,27]. The pile-up is likely promoted by the mechanical mismatch between the soft $Ta_2O_5$ film and the stiff Si substrate. Conversely, the monolithic $SiO_2$ film (Fig. 1h) demonstrated brittle fracture with well-defined radial cracks emanating from the indentation vertices, reflecting its low fracture toughness. The sink-in observed in $SiO_2$ indent, contrasting with the pile-up behaviour exhibited by all other films, indicates densification occurring beneath the indenter, as discussed in detail below.

Multilayers with finer bilayer spacings (4-40 nm; Fig. 1b-e) showed enhanced plasticity and crack resistance relative to $SiO_2$. As bilayer spacing increased to 200-334 nm (Fig. 1f-g), the deformation behaviour transitioned towards that of brittle $SiO_2$, with localised chipping and incipient crack formation. This degradation correlates with reduced interface density, which diminishes interface-mediated toughening mechanisms.

Furthermore, to assess the films' resistance to fracture under severe contact loading, we deliberately made indents of depths exceeding the film thickness, h/t > 1. Fig. S2 shows the remarkable crack resistance offered by the 10- 40 nm bilayer thickness films. Subsequent S/TEM investigation of indent cross-sections explains the observed exceptional plasticity.

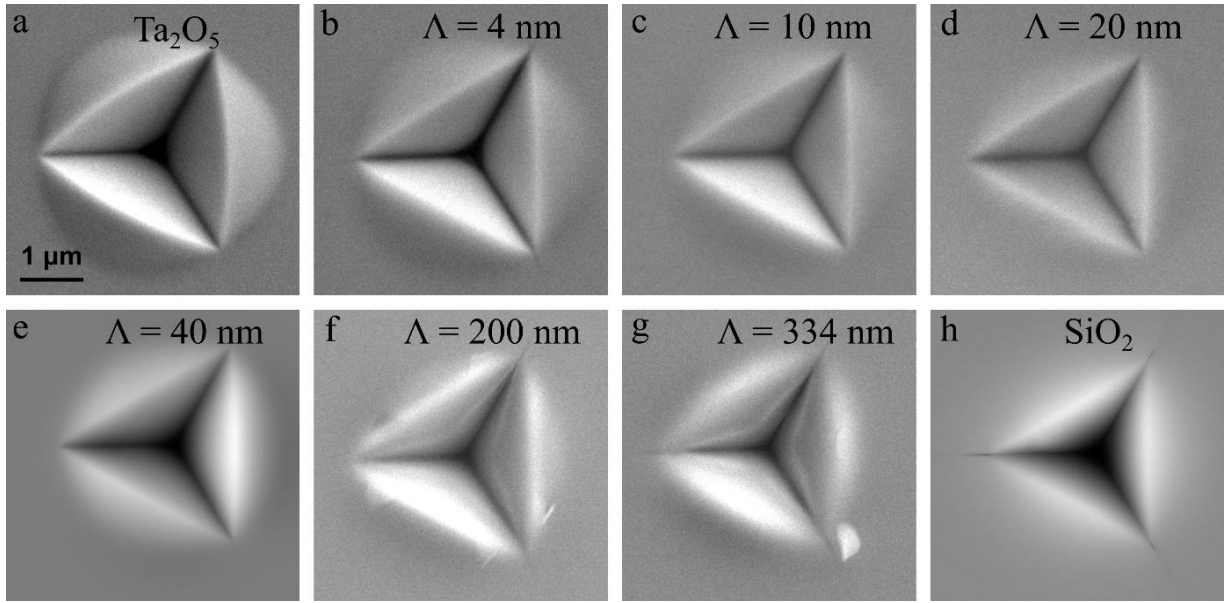

Fig. 1: SEM micrographs of 35 mN nanoindentation impressions on $Ta_2O_5$/$SiO_2$ multilayers with varying bilayer architectures. (a, h) Monolithic $Ta_2O_5$ and $SiO_2$ films exhibit significant pile-up and radial cracking, respectively. (b-g) Multilayers with bilayer spacings of 4 nm, 10 nm, 20 nm, 40 nm, 200 nm, and 334 nm demonstrate a gradual shift from pile-up towards chipping and cracking.

Fig. 2 presents the hardness values for multilayers with bilayer thicknesses ranging from 2 nm to 334 nm, alongside the monolithic films. The monolithic $SiO_2$ film exhibits the highest hardness values at approximately 9.4 GPa, substantially exceeding all multilayer configurations. However, as was evident from the SEM images of the indents, the introduction of a multilayer architecture alters their mechanical response. The multilayer with 334 nm bilayer thickness exhibits hardness values closely matching those predicted by the rule of mixtures, indicating minimal interface effects. Hardness decreases systematically with bilayers, dropping from ~7.7 GPa at 167 nm to 5.5 GPa at 2 nm. This behaviour contrasts markedly with crystalline multilayer systems, which strengthen with interfaces [28,29]. Although pile-up can cause hardness values to be overestimated [2,25], our focus is on relative hardness. Since pile-up is more pronounced in the softer films, the hardness trends shown in Fig. 2 remain reliable and may even be stronger when considering this effect. The insets in Fig. 2 show STEM cross-sectional images beneath an indentation, revealing vertical shear bands with an inter-band spacing of $12 \pm 3$ nm in the 4 nm bilayer film, demonstrating the distributed nature of strain accommodation. The 40 nm bilayer film exhibits wider incipient shear band spacing of $60 \pm 10$ nm accompanied by enhanced interlayer mixing.

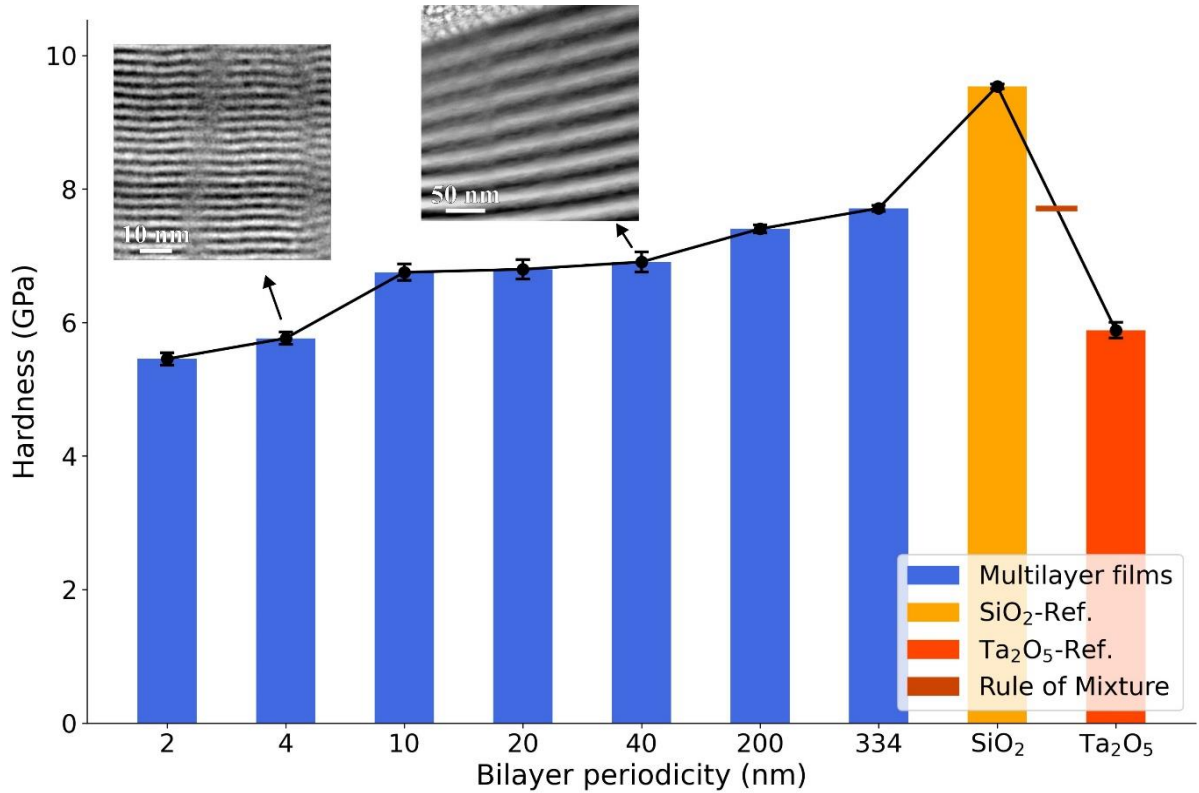

Fig. 2: Hardness of multilayer films as a function of individual layer thickness. Reference values for monolithic $SiO_2$ and $Ta_2O_5$ films and the rule of mixtures prediction are shown for comparison. The insets show STEM cross-sectional images revealing shear bands and interlayer mixing that accommodate deformation.

Fig. 3a shows a BF TEM cross-section of a nanoindent in monolithic $SiO_2$, revealing the shear band propagating through the film thickness. The localised deformation demonstrates the characteristic failure mode of silica under nanoindentation. We observe crystalline-to-amorphous phase transformation in the single-crystal silicon substrate where shear band-substrate interacts. Diffraction patterns in Fig. 3 confirm that directly beneath the indent the deformed substrate (yellow circle) is amorphous. The undeformed (blue circle) regions away from the shear-band-substrate interaction zone retain their single crystalline nature. Similar deformation induced amorphization of Si has been reported earlier [30,31].

On the other hand, the bright region within the shear band is due to its crystallization. HRTEM imaging of the bright region reveals 2 – 5 nm sized small grains, indicative of deformation-induced crystallization. This phenomenon has been previously observed in metallic glasses subjected to severe plastic deformation. Chen et al. demonstrated that such localised shear bands in aluminium-based amorphous alloys form due to local atomic rearrangements in the region of high plastic strain [32]. Similarly, Kim et al. showed that the primary mechanism for crystallization inside shear bands is flow dilatation - the expansion of atomic structure during deformation that creates conditions similar to the glass transition temperature and dramatically enhances atomic diffusional mobility. This enhanced mobility allows atoms to diffuse rapidly enough to form nanocrystallites, without requiring significant temperature rise or other external factors [33]. Our oxide system exhibits analogous behaviour [34–36]. The shear band in Fig. 3a and Fig. 3e has a darker and brighter contrast, respectively due to its densification. In a uniformly thin FIB-prepared TEM sample, the HAADF-STEM contrast in an amorphous material is directly proportional to the density [37]. The same local enhancement in density was also mapped using SPED (Fig. S3), confirming densification as the prominent deformation mechanism in low density $SiO_2$.

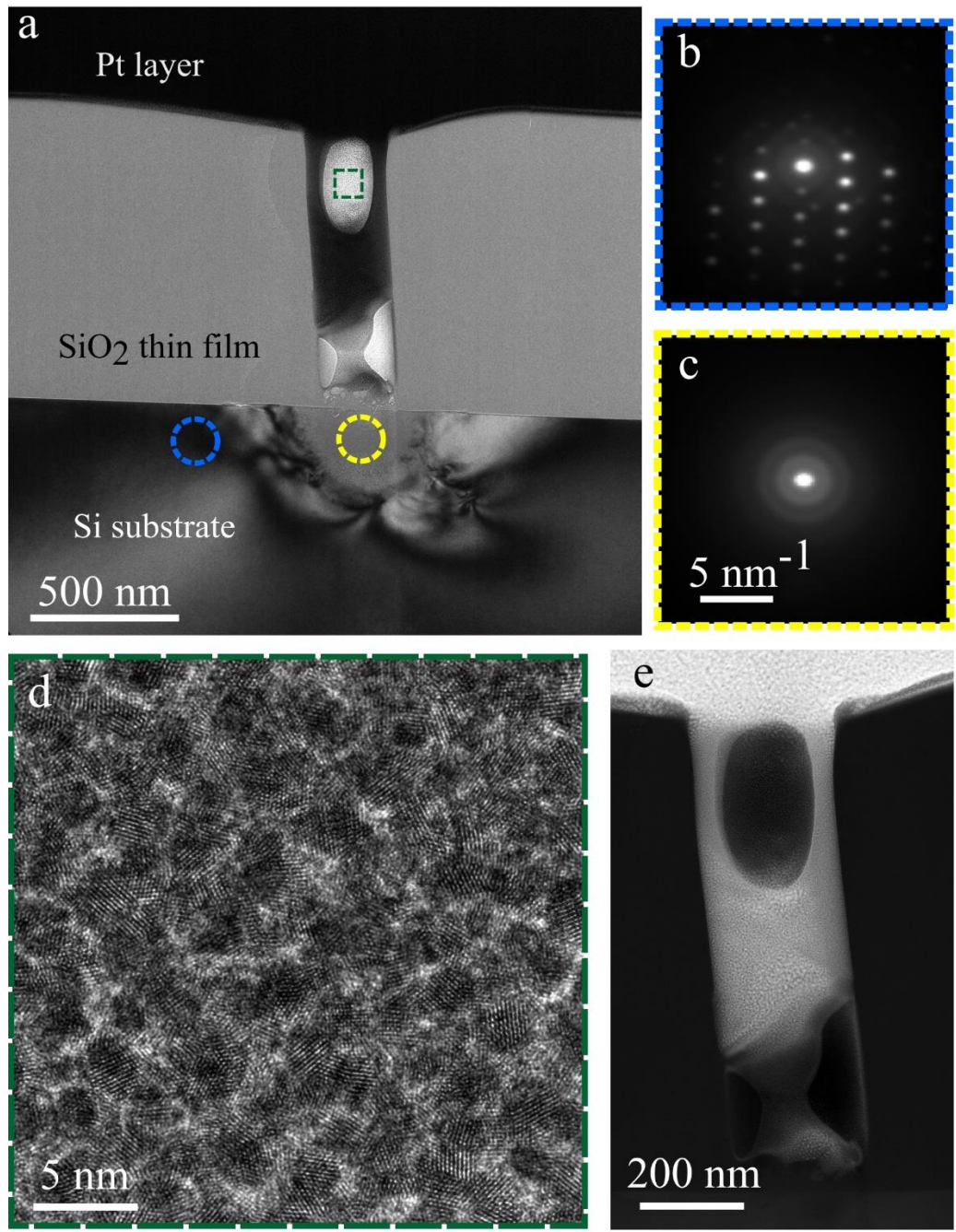

Fig. 3: Deformation-induced crystallization and densification in the amorphous SiO$_2$ thin film and amorphization in the Si substrate during nanoindentation. a) Bright field TEM image shows a single shear band directly beneath the nanoindent impression. Diffraction contrast reveals several other distinct features (highlighted in green, blue and yellow). b-c) Selected area diffraction patterns from apertures marked by blue and yellow circles demonstrate shear-induced amorphization of Si single crystal substrate. d) HRTEM image from the green region reveals crystallization in the shear band, characterised by 3-5 nm nanocrystallites. e) HAADF-STEM image of the same region, highlighting the densification.

In the 4 nm bilayer spacing multilayer (Fig. 4a-g), closely spaced vertical shear bands are observed, with the interlayer spacing decreasing from the original 4 nm to approximately 2.4 nm in the most heavily deformed regions beneath the indent. This uniform compression across all layers effectively distributes the applied strain across multiple deformation pathways. Each interface represents a potential nucleation site for new shear bands while simultaneously impeding the growth of existing ones, leading to enhanced plasticity through interface-mediated deformation mechanisms. Fig. S4 reveals more example images used to measure the spacing between the shear bands. They demonstrate how layer continuity is occasionally disrupted due to shear band propagation in the most deformed regions.

In contrast, the 40 nm bilayer spacing film (Fig. 4h-m) exhibits fewer and more widely spaced constrained shear bands (embryos) that do not coalesce into complete shear bands. The interlayer spacing under the indent is uniformly reduced from 40 nm to approximately 20 nm, demonstrating similar compressive behaviour but with reduced interface density. EDS elemental mapping reveals localised chemical intermixing between $Ta_2O_5$ and $SiO_2$ layers in the deformed regions. Notably, whilst the 4 nm system shows complete layer discontinuity, the 40 nm bilayer thickness film exhibits only chemical intermixing with limited breaking of the layer architecture, indicating a fundamentally different deformation mode. The intensity line profiles across the deformed and undeformed regions clearly distinguish the compositional variations resulting from plastic flow. Crucially, neither multilayer system shows deformation extending to the substrate level, demonstrating the protective effect of the laminated architecture and the confinement of plastic deformation within the multilayer structure. This also suggests that the 10% indentation depth rule does not apply to these films, as also indicated by the plateauing of the depth resolved hardness measurements in Fig. S1. Additionally, the undeformed regions (Fig. 3f and l) demonstrate excellent interfacial quality in the as-deposited film. The individual layer thicknesses exhibit uniformity throughout, while the interlayer interfaces are well-defined and sharp.

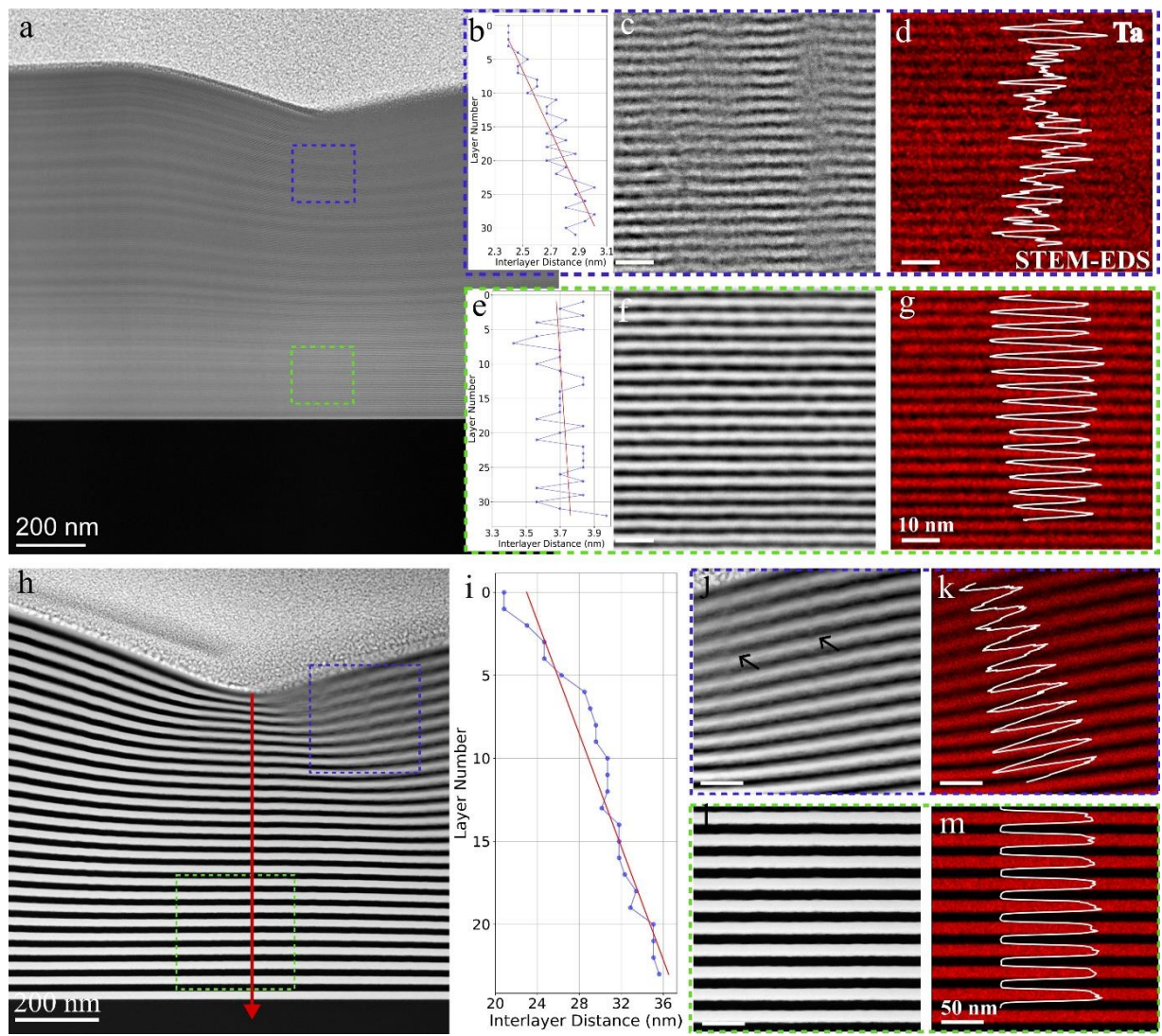

Fig. 4: Cross-sectional microstructural characterization of $Ta_2O_5/SiO_2$ multilayers with the bilayer periods of 4 and 40 nm beneath the indentation region. (a) Low-magnification HAADF-STEM overview

image showing the indented surface region with selected areas (blue and green dashed boxes) for higher magnification analysis. (b) Changes in the bilayer spacing under the indent for the 4 nm film. (c, d) High-magnification HAADF-STEM image and corresponding Ta EDS elemental map of the region directly beneath the indent in the 4 nm film. (e) The bilayer spacing further away from the indent in the 4 nm film appears to be preserved nearly as in the parent film. (f, g) HAADF-STEM images and EDS maps of the region away from the indent. (h) Low-magnification HAADF-STEM overview of the 40 nm bilayer spacing film after indentation, showing the deformed region. (i) Interlayer distance variation profile. (j, k) High-magnification HAADF-STEM image with arrows pointing at the vertical shear bands and corresponding Ta EDS map of the deformed region. (l, m) Undeformed region of the 40 nm film. The scale bars represent 10 nm for high-magnification images of the 4 nm bilayer spacing film, and 50 nm for of the 40 nm bilayer spacing film.

Fig. 5 reveals pronounced structural heterogeneity throughout the indented region. With a 2 nm resolution, SPED was used to measure the densification beneath the indenter in the 40 nm bilayer spacing film as a representative example [38,39]. While the diffraction pattern from undeformed material displays the typical broad, diffuse ring characteristic of an amorphous structure, the deformed region beneath the indent shows a well-defined ring indicating increased density. The normalised radial intensity profiles (Fig. 5e) elucidate this with the deformed region exhibiting a pronounced peak at ~3 $nm^{-1}$.

The density map reveals that the highest density values (red regions) are concentrated immediately beneath the indent centre, with density gradually decreasing with distance from the deformation zone. This gradient parallels the continuous change in bilayer spacing beneath indents (Fig. 4). However, the densification exhibits a more abrupt transition, with a sharper boundary between the severely and moderately densified regions. The map also captures the intrinsic density difference between $SiO_2$ and $Ta_2O_5$. The densification process appears to occur gradually across the deformed volume rather than being concentrated in discrete shear band that accommodates all the deformation as seen in Fig. S3, suggesting that the multilayer structure enables efficient stress transfer between layers. The combination of shear band multiplication and layer densification provides complementary pathways for strain accommodation.

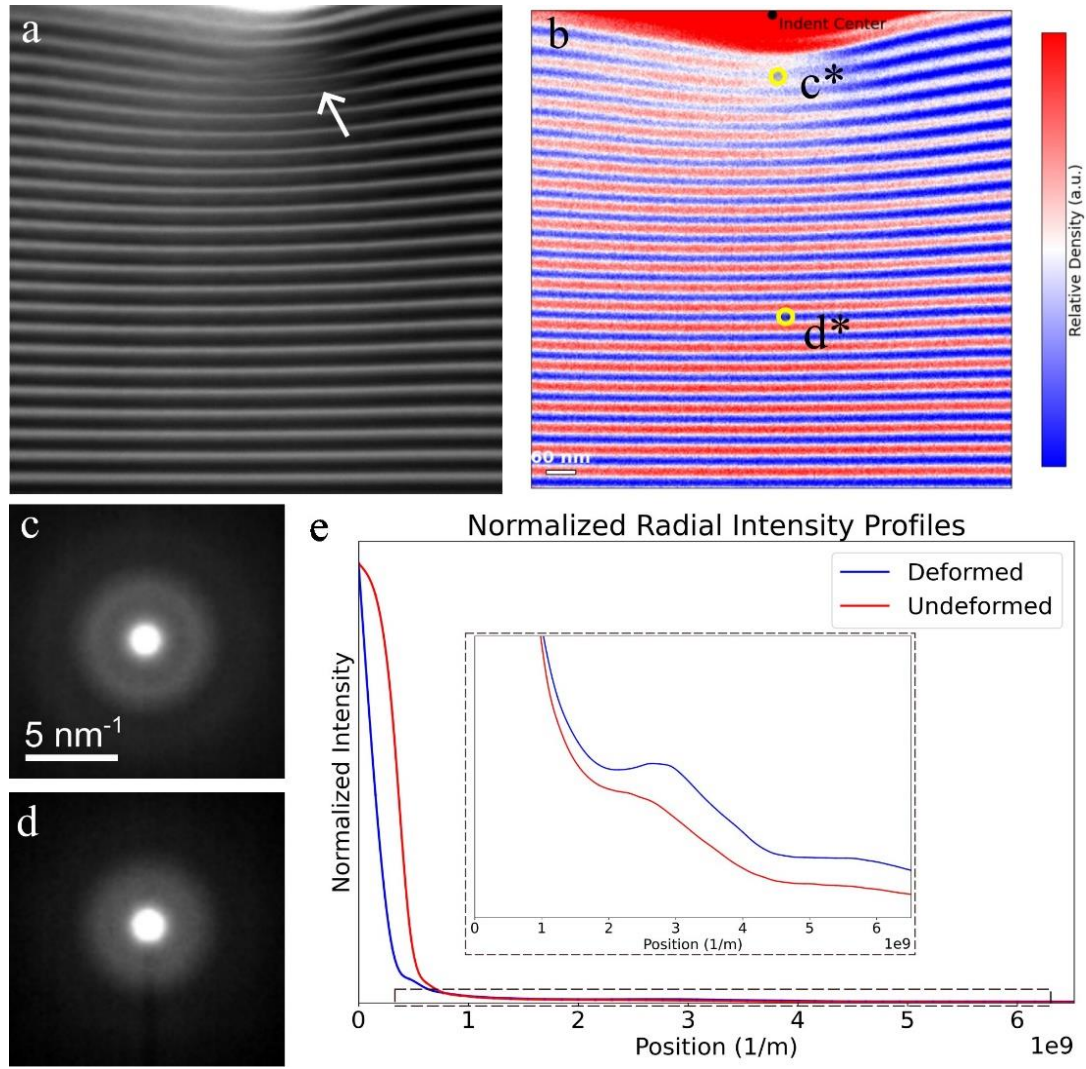

Fig. 5: Diffraction-based density mapping of an indented amorphous film. a) Virtual dark field image of the indented 40 nm bilayer spacing film obtained using the SPED data with an arrow pointed at the disregistery in the nanolaminates. b) Spatial map of relative density calculated from the first moment of radial diffraction profiles, revealing significantly higher density immediately beneath the indent. Diffraction patterns from c) deformed and d) undeformed regions beneath the indent, as indicated in the density map. e) Normalised radial intensity profiles showing the emergence of a peak at 3 nm$^{-1}$ in the deformed region.

Multilayer architecture improves the plasticity and toughness by making the deformation more homogenous. Shear band propagation in amorphous materials is fundamentally dependent on the availability of free volume, which provides the necessary atomic mobility for shear transformation zone activation. Interfaces generally possess excess free volume that can contribute to additional sites for shear band nucleation [40]. However, when individual layers are under a critical thickness, the formation and propagation of mature, localised shear bands can be suppressed. Instead, multiple smaller, "embryonic" shear transformation zones or distributed plastic events occur, leading to a more uniform macroscopic shear flow [30,41,42]. This deformation mechanism transition from highly inhomogeneous flow to relatively more homogeneous deformation. This is reported for both simulations and experiments for metallic and oxide-based A/ANLs [5,10,12,40,42–44]. For example, the deformation homogenization effect was demonstrated through molecular dynamics (MD) simulations on $Mg_{80}Al_{20}/Mg_{20}Al_{80}$ A/ANLs, where strain delocalization was observed as layer thickness decreased below a critical size [44]. Similar enhanced plasticity was reported experimentally using nanoindentation and micropillar compression in Zr/La based A/ANLs [5].

In most reported A/ANLs, Hall-Petch-like strengthening occurs with increasing interface density. Adding interfaces blocks shear bands, the primary carriers of plasticity, thereby strengthening the material. For example, Kuan et al. showed that in ZrCuTi/ PdCuSi system, the propagation of shear bands is arrested due to significant elastic force at the interfaces resulting in increased strength [10]. More recently, Poltronieri et al., pointed out that the enhanced strength in their CuZr-based A/ANLs was due to the local chemical variation at the interface [15]. By contrast, our $Ta_2O_5$/$SiO_2$ A/ANLs show systematic softening with more interfaces.

The intrinsic mechanical properties of the constituent materials provide crucial insight into our findings. $Ta_2O_5$ is inherently ductile. During nanoindentation, $Ta_2O_5$ films characteristically display significant pile-up around indents. In stark contrast, $SiO_2$ exhibits brittle behaviour characterised by densification under mechanical stress (up to 20% volume change) during indentation [34,45]. When these materials are combined to form nanolaminates, their deformation behaviour is modified. The films with 2 or 4 nm bilayer spacing show markedly little pile-up or sink-in. The film shows plasticity due to densification, chemical intermixing and co-deformation due to homogenous shear banding. This behaviour parallels observations in $Al_2O_3$/$Ta_2O_5$ nanolaminates, where the presence of harder $Al_2O_3$ layers reduced pile-up from approximately 7 nm in pure $Ta_2O_5$ to around 4 nm in the nanolaminates [27]. , confirming that harder layers constrain $Ta_2O_5$ flow. Conversely, localized shear banding of $Ta_2O_5$ supresses catastrophic $SiO_2$ cracking. Thus, interfaces simultaneously nucleate and block shear bands, creating a balance of strengthening and softening that dictates overall mechanical response.

In conclusion, $Ta_2O_5$/$SiO_2$ nanolaminates exhibit interface-controlled deformation mechanisms that fundamentally differ from crystalline multilayer systems. The interfaces nucleate and multiply shear bands, promote densification, and confine deformation. The cracking in $SiO_2$ is blocked by $Ta_2O_5$ at nanoscale. Unlike crystalline nanolaminates that show Hall Petch-like strengthening, the oxide A/ANLs represent a distinct class of materials where interfaces enhance toughness by softening.

**Funding**

This work was supported by Innosuisse - Swiss Innovation Agency (Project 110.023 IP-ENG).

**Supplemental figures**

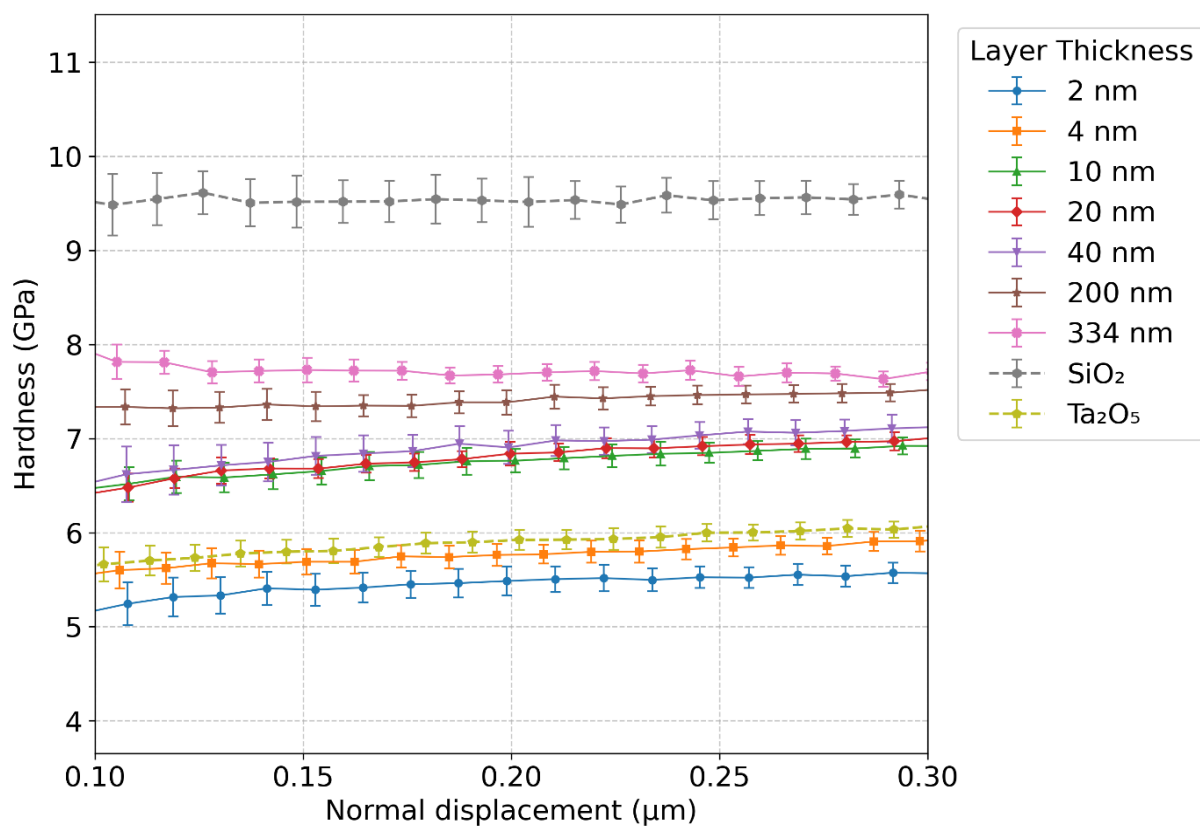

Fig. S1: Nanoindentation hardness of all films plotted again indent depth. Although the indents were deeper than 10% of film thickness criterion [24,25], the depth resolved hardness profiles for all the films with plateauing of the hardness values at increasing depths of indent suggested minimal interaction of the indents with the substrate.

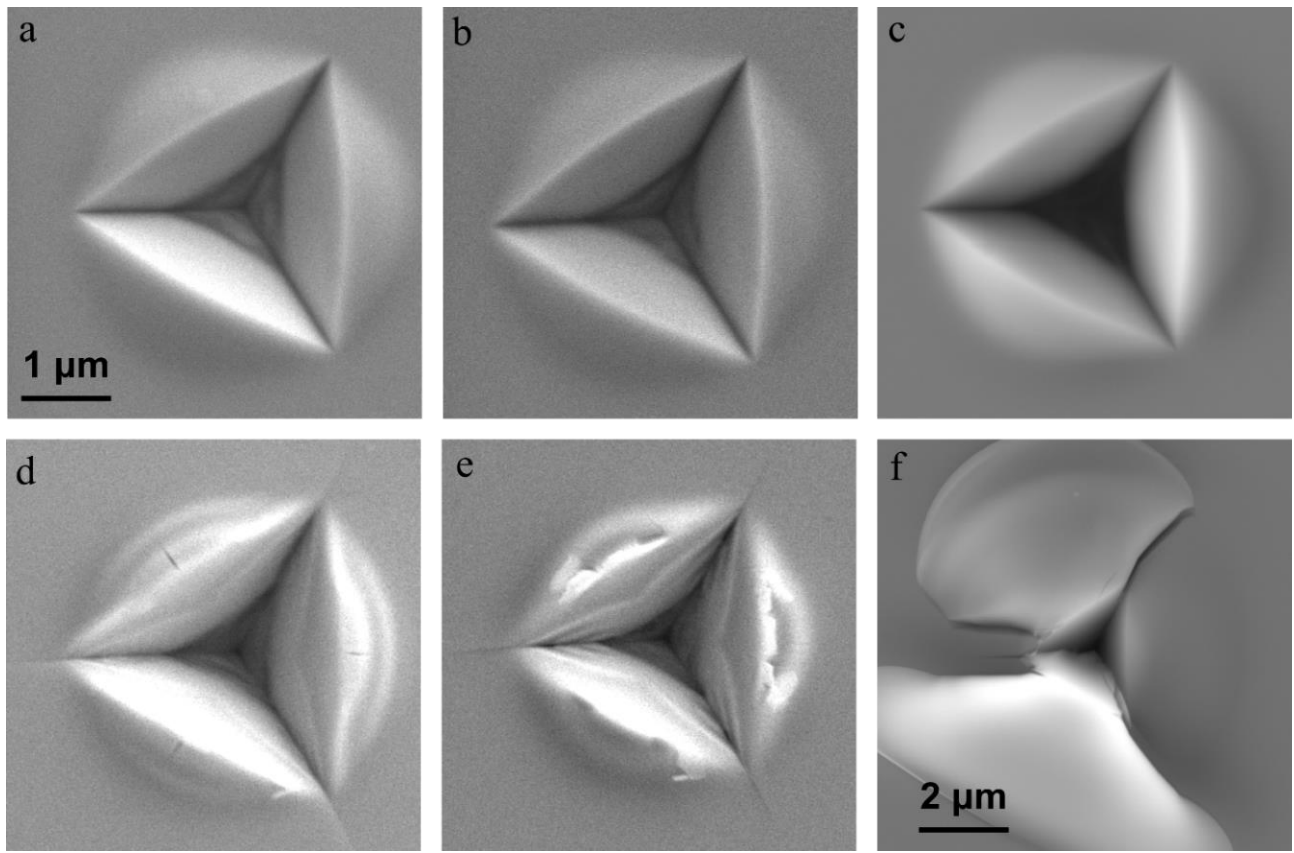

Fig. S2: SEM micrographs of 100 mN nanoindentation impressions on $Ta_2O_5/SiO_2$ multilayer films with varying bilayer architectures with nanoindentation depth larger than the film thickness, extending into the Si substrate. a-c) Multilayers with bilayer periods of 10 nm, 20 nm, 40 nm, respectively, displaying pronounced material pile-up. d, e) Coarser multilayers with bilayer periods of 200 nm and 334 nm transition towards brittle behaviour with evidence of material chipping and crack propagation adjacent to the indentation site. f) $SiO_2$ monolith demonstrating brittle cracking.

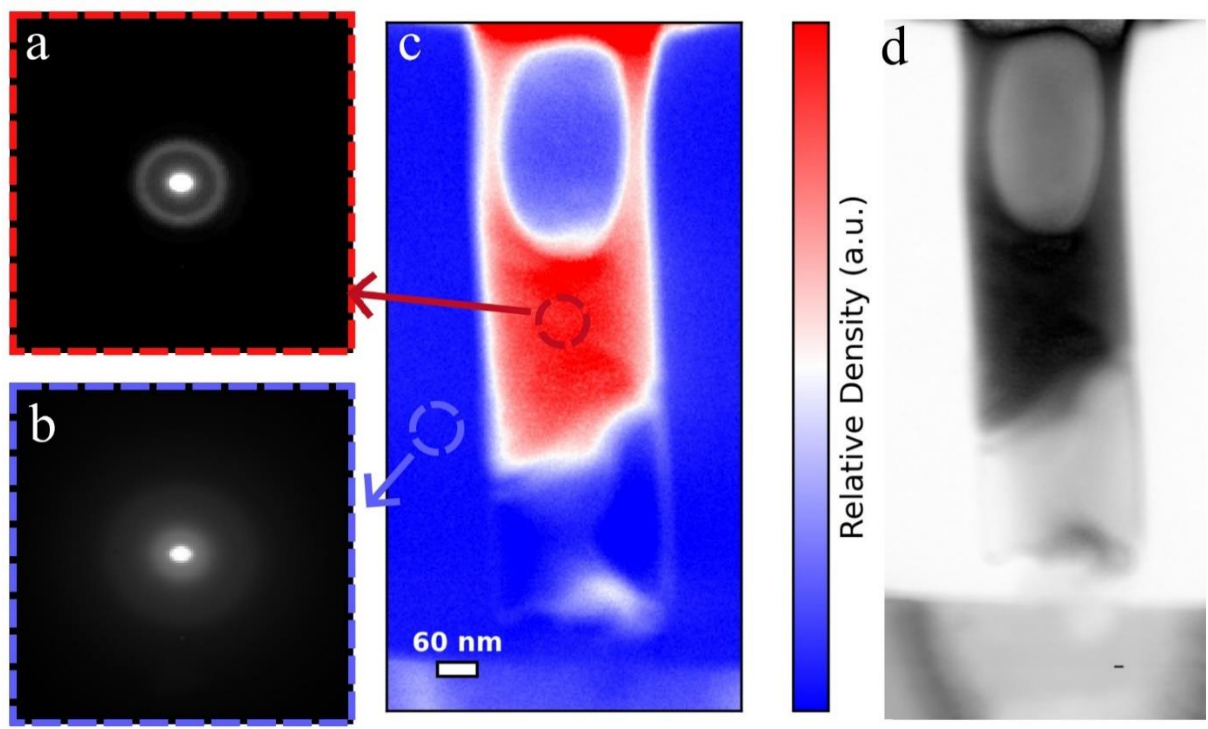

Fig. S3: Relative density distribution across the shear band mapped using SPED. a, b) Diffraction patterns acquired from regions inside and outside the shear band, illustrating the pronounced increase in density within the shear localization zone. The positions of the virtual apertures are indicated by dashed circles. d) Corresponding virtual bright-field image from the same region.

It should be noted that although the crystalline $SiO_2$ region (in the top half of the shear band) is displayed in blue, it is actually denser than the surrounding amorphous film. In this analysis, density is estimated using radial integration of the diffraction intensity as a first-order approximation. However, the method does not account for the relative intensity of individual diffraction spots. As a result, the crystalline region can misleadingly appear less dense, since the azimuthal integration around the transmitted beam is dominated by only a few strong reflections, reducing the first-moment contribution. Therefore, the density map should be interpreted strictly in terms of comparing the shear band region to the undeformed amorphous matrix, while the crystalline region should not be included in the comparison.

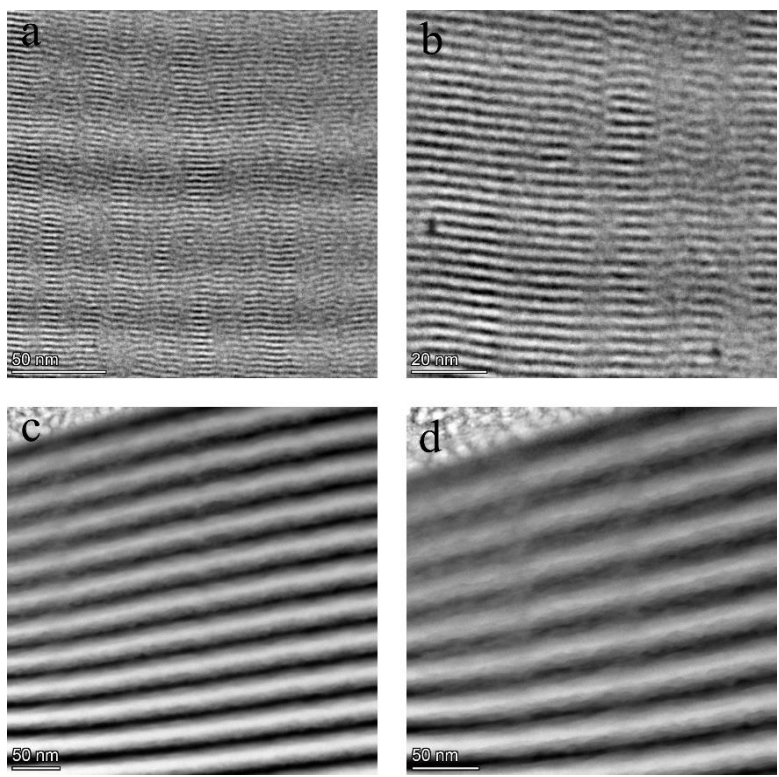

Fig. S4: HAADF-STEM image of the region under the indent in a, b) 4 nm bilayer spacing film and c,d) in the 40 nm bilayer spacing film. These exemplary images were used to evaluate the average shear band spacing in both the films.